\documentclass[final]{elsarticle}
\usepackage{graphicx}
\usepackage{amsmath}
\usepackage{lineno,hyperref}
\usepackage{soul} 
\usepackage{setspace}

\modulolinenumbers[50]

\journal{Journal of CEE}

\begin{document}

\begin{frontmatter}

\title{The link to the formal publication is via\\
	{\small \url{http://dx.doi.org/10.1016/j.compeleceng.2017.06.008}}\\ .\\ An Analysis Framework for Hardware and Software Implementations with Applications from Cryptography}

\author[IssamAddress]{Issam Damaj\corref{mycorrespondingauthor}}
\author[SafaaAddress]{Safaa Kasbah}
\cortext[mycorrespondingauthor]{Corresponding Author}

\address[IssamAddress]{Electrical and Computer Engineering Department, American University of Kuwait, Salmiya, Kuwait, idamaj@auk.edu.kw}

\address[SafaaAddress]{Computer Science and Mathematics Department, Lebanese American University, Beirut, Lebanon, safaakasbah@gmail.com}

\begin{abstract}
With the richness of present-day hardware architectures, tightening the synergy between hardware and software has attracted a great attention. The interest in unified approaches paved the way for newborn frameworks that target hardware and software co-design. This paper confirms that a unified statistical framework can successfully classify algorithms based on a combination of the heterogeneous characteristics of their hardware and software implementations. The proposed framework produces customizable indicators for any hybridization of processing systems and can be contextualized for any area of application. The framework is used to develop the Lightness Indicator System (\textit{LIS}) as a case-study that targets a set of cryptographic algorithms that are known in the literature to be tiny and light. The \textit{LIS} targets state-of-the-art multi-core processors and high-end Field Programmable Gate Arrays (\textit{FPGAs}). The presented work includes a generic benchmark model that aids the clear presentation of the framework and extensive performance analysis and evaluation. 

\end{abstract}


\end{frontmatter}


\section{Introduction}

With the advancements in high-performance computing, algorithms have a wide range of efficient implementation options. Current computers can be equipped with multi-core processors, Graphics Processing Units (\textit{GPUs}), and high-end programmable devices, such as, \textit{FPGAs}. The variety of processing options are supported by a wealth of co-design tools that facilitates hardware and software implementations \cite{JD072,KDH08}. Nevertheless, several questions remain on what algorithm is the best to suite an implementation option, and vice-versa. How would an algorithm perform within hybrid processing systems, and how to make an evaluation based on heterogeneous performance measurements?

The core of any performance measurement includes measures, metrics, and indicators. Indicators are defined as qualitative or quantitative factors, or variables that provide simple and reliable means to measure achievement. A qualitative performance indicator is a descriptive characteristic, an opinion, a property or a trait. However, a quantitative performance indicator is a specific numerical measurements resulted by counting, adding, averaging numbers or other computations \cite{asbjorn1995benchmarking}. Qualitative and quantitative measurements can be combined to define measurement frameworks and benchmarks \cite{damajsustainability}. There is a large number of hardware and software benchmarks in the literature. Yet, limited research work is reported to address developing analysis frameworks for heterogeneous hardware and software implementations. 

In this paper, we present a statistical analysis framework for performance profiling of related algorithms running under different hardware and software subsystems. The framework comprises criteria, indicators, and measurements obtained from heterogeneous sources. The measurements are statistically combined to produce indicators that capture the algorithmic, software, and hardware characteristics of the assessed algorithms. The developed framework enables the deep and thorough reasoning about each hardware and software subsystem, and combines heterogeneous characteristics to provide overall ratings, rankings, and classifications. The proposed framework is customizable for any hybridization of processing systems and can target any model of computation or area of application. 

The paper includes the development of a generic benchmark model that serves as a specification pattern of analysis and evaluation frameworks. The model captures the activities, resources, implementation, mathematical formulation, and intended measurements of an analysis framework or a benchmark. The developed model can be used to describe any benchmark with simplicity and clarity. The model is adopted to present the proposed analysis framework.   

To validate the proposed framework in application, a case-study is carried out with application from cryptography. The case-study enables the development of the\textit{LIS} with its bouquet of statistical indicators. The \textit{LIS} formulates the proposed framework within the context of lightweight cryptographic algorithms. The proposed performance analysis classifies the investigated algorithms into a combination of their mathematical, software, and hardware characteristics. The two main targeted high performance computing devices are multi-core processors for software implementations and \textit{FPGAs} for hardware implementations.

The rest of the paper is organized so that Section~\ref{Benchmarks} surveys the literature. In Section ~\ref{Research Objectives}, the motivation, research questions, and the paper contribution are presented. In Section ~\ref{The Generic Model and the Analysis Framework}, the generic benchmark model and the analysis framework are presented. Section ~\ref{The Application of the LIS System to Cryptographic Algorithms} introduces the \textit{LIS} according to the generic model.  A thorough performance analysis and evaluation is presented in Section ~\ref{Performance Analysis and Evaluation}. Section ~\ref{Conclusion} concludes the paper and sets the ground for future work.

\section{Related Work} \label{Benchmarks}
\subsection{Benchmarks}

 Benchmarks are widely addressed in the literature. Famous benchmarks include Whetstone, LINPAC, Dhrystone, Standard Performance Evaluation Corporation (SPEC), etc \cite{Linpack2011,weicker1984dhrystone,henning2000spec}. Several developments of embedded systems Benchmarks are lead by the Embedded Microprocessor Benchmark Consortium (\textit{EEMBC}). \textit{EEMBC} helps system designers in selecting the optimal processors, smartphones, tablets, and networking appliances. \textit{EEMBC} mainly targets embedded system's hardware and software. \textit{EEMBC} organizes its benchmark suites targeting automotive, digital media, multi-core processors, networking, automation, signal processing, hand-held devices, and browsers. The benchmarks developed by \textit{EEMBC} include \textit{AutoBench}, \textit{BrowsingBench}, \textit{AndBench}, and \textit{MiBench} \cite{EEMBC2014}. Cryptography benchmarks are designed to measure the performance of different cryptographic algorithms running under different systems, such as, \textit{GPUs} or other processors. Rukhin et al. in \cite{rukhin2001statistical} presents a statistical test suite for random and pseudorandom number generators for cryptographic applications. Yue et al. in \cite{yue2006npcryptbench} presents a cryptographic benchmark suite for network processors (NPCryptBench).

\subsection{Hardware/Software Co-Design Evaluation Frameworks}

Performance analysis and evaluation within hardware/software (HW/SW) co-design investigations are usually based on a variety of metrics. Besides standard metrics, such as execution time, maximum frequency, throughput, hardware resource utilization, power consumption, etc., several metrics are identified within the context of application. Jain-Mendon and Sass in \cite{Jain-Mendon2014873} proposed a HW/SW co-design approach for implementing sparse matrix vector multiplication on \textit{FPGAs}. Within the context of application, the authors evaluated their approach by analyzing the hardware and software implementations in terms of the speed of processing floating point operations, bandwidth efficiency, data block size, communication time, etc. Lumbiarres-Lopez et al. \cite{LumbiarresLopez2016324} implemented, within a co-design environment, a countermeasure against side-channel analysis attacks. The used application-specific metrics comprise the difference in change of input current over time and correlations between data and power consumption. All the aforementioned investigations employed the standard co-design metrics. In \cite{Park20131578}, the performance of block tridiagonal solvers was evaluated under heterogeneous multi-core processors and \textit{GPUs}. The evaluation was mainly based on analyzing memory performance and measuring the total execution times of different scenarios.


The standard metrics of co-design applies to partitioned hardware and software implementations. The focus in partitioned implementation is the analysis and evaluation of the developed subsystems with an aim to find the best possible partitioning strategy. Wu et al. \cite{Shi20147} studied the performance and algorithmic aspects of a proposed heuristic partitioning algorithm. The produced implementations were analyzed with-respect-to execution time, resource utilization, and the attained solution quality as related to the smallest possible error. In \cite{Jemai2015259}, Jemai and Ouni proposed a partitioning strategy based on control data graphs. The partitioning algorithm was deployed within three different case-studies. The metrics analyzed across the three studies comprised the number of partitions, software execution time, hardware resource utilization, software resource utilization, etc. The authors adopted a pattern chart that summarizes the performance analysis results and aids the evaluational of the algorithm. 

Nevertheless the evaluation approaches for HW/SW co-design considered various aspects, limited attempts were made to combine multiple measurements and characteristics in unified indicators. Spacy et al. in \cite{Spacey2009159} investigated the automatic quantification of acceleration opportunities for programs across a wide range of heterogeneous architectures. The investigation focused on allowing designers to identify promising implementation platforms before investing in a particular HW/SW co-design and a specific partitioning scenario. The authors unified many hardware and software characteristics into a single execution time estimate. The incorporated hardware characteristics included cycle time, number of parallel execution units, execution efficiency, bus latency, bus width, and hardware size capacity. The employed software characteristics included the execution time, the number of parallel execution slots, program code unit iterations, control flows, data flows, and size of codes. Additional composite indicators were developed to calculate speedup factors. The combined characteristics were used to calculate co-design performance estimates and evaluating opportunities for hardware acceleration. 

\section{Research Objectives} \label{Research Objectives}
The modern trend in computer systems is clearly in the direction of further hybridization using high-end co-processing systems. Hybrid systems are mainly studied within HW/SW co-design and co-analysis frameworks. To increase the effectiveness of co-analysis and accordingly co-design, the following research opportunities are highlighted:

{\setstretch{1.0}
\begin{itemize}
	\item The identification of commonly-used metrics in HW/SW co-design
	\item Addressing the need for the contextualization of application-specific metrics, properties, and key indicators in HW/SW co-design applications
	\item Evaluating implementations - given the heterogeneous characteristics of the targeted systems
	\item The identification of optimized combinations of hardware and/or software implementations based on co-analysis
	\item The limited attempts in the literature to combine multiple measurements and characteristics in unified indicators that can rank, rate, classify, and evaluate algorithms for hardware and software implementations
	\item The limited work in the literature to develop co-analysis frameworks that target cryptographic algorithms  
\end{itemize}}

The research objective of this paper is mainly to develop a statistical framework that can combine heterogeneous characteristics of algorithms and their implementations in hardware and software. The framework aims at being portable across different hardware and software systems, customizable, scalable, and able to target any area of application. In addition, the framework aids the composition of a bouquet of indicators to capture specific desirable properties and enable classifying, ranking, rating, and evaluating algorithms and implementations. In addition, the objectives include the follows:

{\setstretch{1.0}
\begin{itemize}
	\item Provide a generic benchmark model that serves as a specification pattern of analysis and evaluation frameworks. The developed model aims at being clear, simple, and highly reusable. The model is used to present the developed analysis framework.
	
	\item Validate the proposed framework by developing the \textit{Lightness Indicator System} for cryptographic ciphers.

	\item Perform a thorough analysis and evaluation based on the \textit{LIS} system for a set of cryptographic ciphers.
	
	\item Studying the integration of the developed framework within and Integrated Development Environment (\textit{IDE}) that can connect to various hardware and software implementation and analysis tools.
	
\end{itemize}}

The paper includes a thorough evaluation of the framework and a discussion on its usefulness.

\section{The Generic Model and The Analysis Framework} \label{The Generic Model and the Analysis Framework}

\subsection{The Generic Benchmark Model}
The proposed generic model diagrams the continuum of important elements of benchmarks and analysis frameworks. The generic model defines the goal, inputs, activities, output, outcomes, and the desired performance profile of a benchmark. The model captures the relationships among the resources, implementation, mathematical formulation, and the obtained results. Moreover, it standardizes the evaluation process that can be applied to any benchmark. The proposed model consists of the following six elements; the model elements are diagrammed in Figure \ref{fig:GSModel}:

\begin{figure}
   \caption{The six elements diagram of the Generic Benchmark\textit{} Model with the required specifications}
   \label{fig:GSModel}
   \centering
        \includegraphics[width=0.57\linewidth]{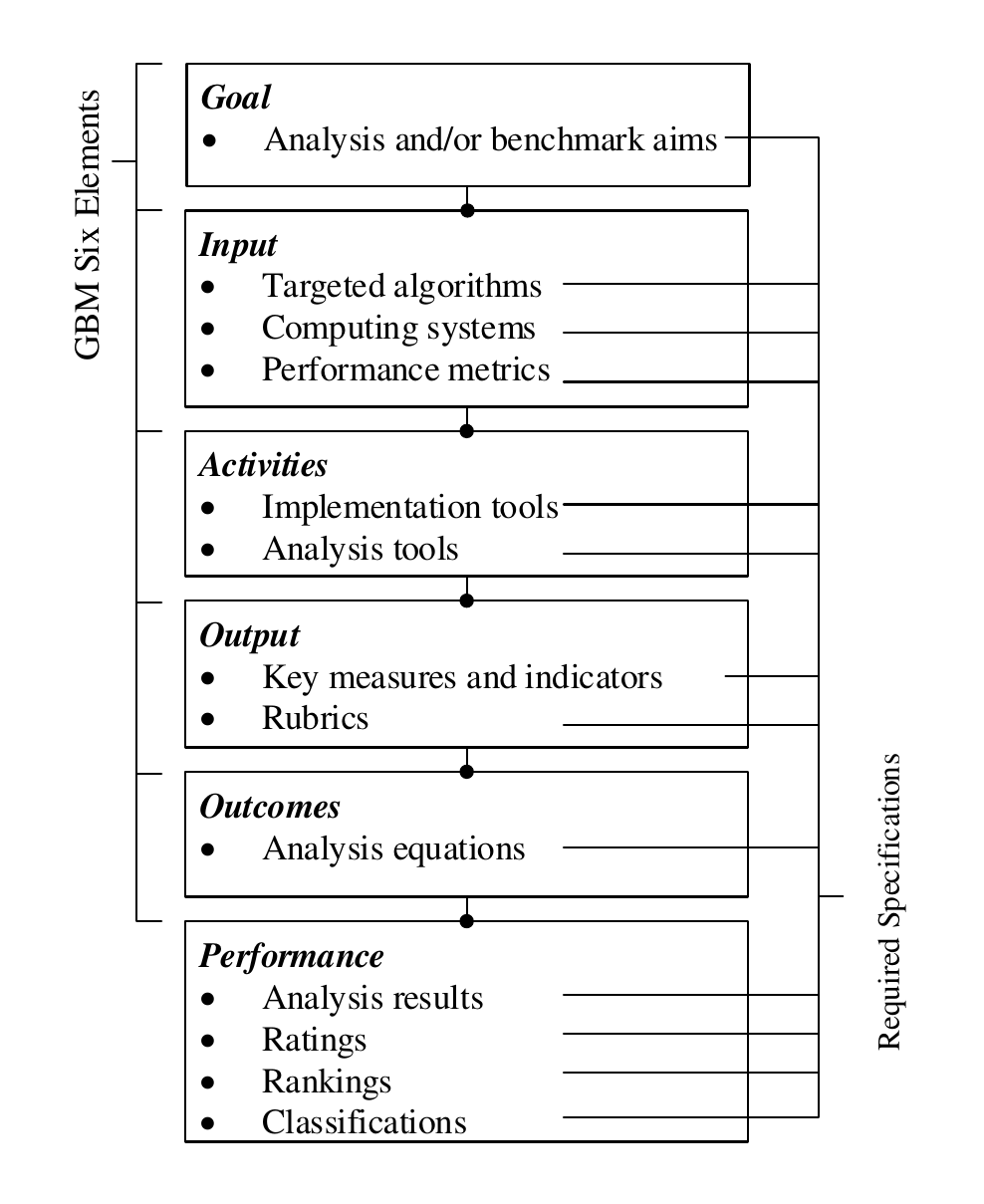}
\end{figure}

{\setstretch{1.0}
\begin{enumerate}
  \item \textbf{Goal}: is the definition of the aim of the benchmark, or the analysis framework, and what does it mainly provide.

  \item \textbf{Input}: is the identification of the algorithms under study, implementation environments, reference algorithm, performance metrics, etc.

  \item \textbf{Activities}: are the implementations of the algorithms under the identified environments and collection of results.

  \item \textbf{Output}: is the formulation of the key indicators and development of their rubrics - if any.

  \item \textbf{Outcomes}: are the formulations of the statistical assessment as combinations of the Output.

  \item \textbf{Performance}: is the application of the developed assessment framework to profile and classify algorithms according to the obtained results.
\end{enumerate}}

The proposed model provides a generic profiling pattern than can be used for any benchmark or analysis framework.

\subsection{The Analysis Framework} \label{The Analysis Framework}
The proposed analysis framework classifies the heterogeneous sources of measurements into analysis profiles (APs), such as, general algorithmic, hardware, and/or software. The development of each profile includes the identification of a set of Key Indicators (\textit{KIs}). The indicators are the most extensive part of the measurement framework and should be carefully developed within the context of application. The measurements associated with the identified indicators may include quantities, scores from scale-rubrics, etc. The measured indicators are then each divided by a measurement from a reference algorithm for normalization and for producing performance ratios. Accordingly, Combined Measurement Indicators (\textit{CMIs}) are calculated using the Geometric Mean of \textit{KI} ratios. The generic equation of \textit{CMIs} is as follows:\\

$CMI = \sqrt[n]{ratio_{1} \times ratio_{2}\times... ratio_{n}}$\\

Where $ratio_{i} =  \dfrac{KI_{i.j}}{KI_{i.j}^{ref}}$\\ 

$KI_{i.j}$ is the $i^{th}$ \textit{KI} of the $j^{th}$ \textit{AP},


and $KI_{i.j}^{ref}$ is the reference measurement of the indicator $KI_{i.j}$\\

The Geometric Mean is used, in the \textit{CMI} equation, as it is able to measure the central tendency of data values that are obtained from ratios. Using the Geometric Mean insures the following two important properties \cite{bullen2003handbook,denscombe2014good,hennessy2011computer}:

{\setstretch{1.0}
\begin{itemize}
	\item Geometric Mean of the ratios is the same as the ratio of Geometric Means.
	\item Ratio of the Geometric Means is equal to the Geometric Mean of performance ratios; which implies that when comparing two different implementations' performance, the choice of the reference implementation is irrelevant \cite{hennessy2011computer}.
\end{itemize}}

The developed statistical framework can be applied in different areas of applications and using different \textit{APs}. The application includes contextualizing the \textit{KIs} of every \textit{AP} according to the characteristics of the targeted area. 

\section{The Application of the LIS System to Cryptographic Algorithms} \label{The Application of the LIS System to Cryptographic Algorithms} In the following sections, we present the \textit{LIS} based on the generic model.

\subsection{Goal}

The \textit{LIS} is a HW/SW statistical framework that provides performance profiling for lightweight cryptographic algorithms running under different hardware and software systems. The \textit{LIS} combines a bouquet of performance metrics that include speed, algorithmic complexity, memory efficiency, algorithmic strength, hardware size, etc.

\subsection{Input}

The input identifies the targeted algorithms, computing systems, and the performance metrics. The \textit{LIS} targets a set of cryptographic algorithms that are specific for applications running on low resources. The selected ciphers are classified, in the literature, as tiny, small, lightweight, or ultra-lightweight ciphers. The targeted ciphers are \textit{Skipjack}, \textit{3-WAY}, \textit{XTEA}, \textit{KATAN} and \textit{KTANTAN}, and \textit{Hight}. The reference cipher is the \textit{AES} \cite{daemen2002design}.


The two targeted high performance computing devices are the \textit{Dell Precision T7500} with its dual quad-core \textit{Xeon processor} and \textit{24 GB} of \textit{RAM}. The targeted \textit{FPGA} is \textit{Altera} \textit{Stratix-IV}.


The identified performance metrics of the \textit{LIS} are classified into general algorithmic, hardware, and software profiles. The general algorithmic profile includes the complexity of the algorithms and their security strength. Within the multi-core environment, the software profile includes execution times, clocks per instruction, throughput, and a cache analysis. Within the \textit{FPGA} environment, the hardware profile includes the resource utilization, propagation delay, throughput, power consumption, etc.

\subsection{Activities}

The activities includes software implementations under \textit{C} and hardware implementations under \textit{VHDL}. The Software tools used for software and hardware implementations and profiling are \textit{Quartus}, \textit{ModelSim}, and \textit{Intel VTune Amplifier} under Visual Studio.

\subsection{Output}
The outputs of the analysis framework are measures and indicators. The measures are the general algorithmic, hardware, and software profiles. The indicators of the general algorithmic profile intend to capture the complexity and ciphering strength of the algorithm; the \textit{KIs} include the following in bold:

{\setstretch{1.0}
\begin{itemize}
  \item \textbf{Algorithm Complexity} (\textit{AC}): Asymptotic complexity analysis using the Big-$O$, small-$\omega$, and $\varTheta$ notations
  \item Cipher Strength: based on \textbf{Key Size} (\textit{KS}), \textbf{Number of Rounds}(\textit{NR}), and the text \textbf{Block Size} (\textit{BS})
\end{itemize}}

Complexity analysis of algorithms is the determination of the amount of resources necessary to execute them. To analyze the complexity of studied algorithms, we study their asymptotic behavior. The asymptotic behavior classifies algorithms according to their rate of growth with respect to the increase in input size. The following standard complexity analysis classification is adopted \cite{cormen2001introduction}: 

{\setstretch{1.0}
\begin{itemize}
	\item $O(f(n))$: The rate of growth of an algorithm is asymptotically no worse than the function $f(n)$ but can be equal to it. 
	\item $\omega(f(n))$: The rate of growth of an algorithm is asymptotically no better than the function $f(n)$. 
	\item $\varTheta(f(n))$: The rate of growth of an algorithm is asymptotically equal to the function $f(n)$. 
\end{itemize}}

\noindent Here, $n$ is the size of input.\\

To facilitate the assessment of the studied ciphers, a rubric is created. The rubric scale points are logarithmic low (\textit{LL}), logarithmic high (\textit{LH}), Linear (L), Almost Quadratic (\textit{AQ}), and Higher than Quadratic (\textit{HQ}). For instance, \textit{LL} describes the case when the complexity is asymptotically no worse than $logn$ but can be equal to it; such a complexity is formulated as $O(logn)$. The complete description of the rubric is shown in Table \ref{tab:CARubric}. In preparation for the statistical formulation, we map this qualitative properties onto quantities. For every point in the scale, we map it onto a fixed number. Hence, each point in the scale is mapped onto the values 20\%, 40\%, 60\%, 80\%, and 100\%.


\begin {table*}
\small
\caption {The rubric of the Complexity Analysis indicator} \label{tab:CARubric}
\centering{
\begin{tabular}{|p{1.8cm}|p{1.9cm}|p{1.9cm}|p{1cm}|p{1.6cm}|p{1.5cm}|}
\hline
\multicolumn{1}{|l|}{\textbf{General}} & \multicolumn{5}{|c|}{\textbf{Scale}}\\
\hline
\textbf{Indicator}& \textbf{Logarithmic Low}& \textbf{Logarithmic High}& \textbf{Linear}& \textbf{Almost Quadratic}& \textbf{Higher than Quadratic}\\
\hline
\textbf{Complexity Analysis}& $O(logn)$ & $\omega(logn)$ but better than Linear&$\varTheta(n)$&$O(n^{2})$ but worse than Linear & $\omega(n^{2})$\\
\hline
\end{tabular}}
\end{table*}

Cipher Strength is an assessment of the algorithm based on a variety of aspects that can include Key Size, the Number of Rounds, and the Block Size \cite{jorstad1997cryptographic}. Key size or key length is the size measured in bits of the key used in cryptographic algorithms. The security of the cryptographic algorithms is function of the length of its key. For some algorithms, such as those targeted in this investigation, the longer the key, the more resistant is the algorithm \cite{menezes2010handbook}. However, in the broader context, the relation between key lengths and security could be more delicate \cite{lenstra2004key}. For example, key sizes of 80, 160, and 1024 bits, nevertheless different, they imply comparable security when 80 is for a symmetric cipher, 160 is for a hash length, and 1024 is for RSA modulus \cite{lenstra2004key}. In addition, Elliptic Curve Cryptography (ECC) is famous for its strength that can be attained at relatively small key sizes. For instance, a comparable security level can be achieved using RSA with a key size of 15360 bits and ECC with a key size of only 512 bits \cite{maletsky2015rsa}. Investigations relating the level of security, and the strength of the algorithm, to the key size are given wide and careful attention in the literature \cite{lenstra2004key,maletsky2015rsa,Keylength2017}. The most recent standardized key size requirements for security are published at \cite{Keylength2017}.

Furthermore, block ciphers transform a plain-text block of several bits into an encrypted block. The block size cannot be too short in order to secure the cryptographic scheme. In other words, the larger the block size is, the greater the cipher strength \cite{menezes2010handbook}. In addition, rounds are important to the strength of ciphers; a single round is usually responsible for mixes, permutations, substitutions, and shifts in the text being encrypted. Mostly, more rounds lead to greater confusion and diffusion and hence stronger security. Indeed, indicators like the Key Size, Number of Rounds, and Block Size should be carefully adopted and specified within the scope of the targeted cryptographic algorithms. The proposed indicators are not necessarily applicable to all cryptographic algorithms.

The software profile includes the following indicators\cite{PattHenn2013}:

{\setstretch{1.0}
\begin{itemize}
  \item \textbf{Execution Time} (\textit{ET}): the time between the start and the completion of a task.
  \item \textbf{Throughput} (\textit{TH}): the total amount of work done in a given time .
  \item \textbf{Clock Cycle per Instruction} (\textit{CPI}): the average number of clock cycles each instruction takes to execute.
  \item \textbf{Cache Miss Ratio} (\textit{CMR}): the ratio of memory accesses cache miss.
\end{itemize}}

The hardware profile includes \textit{ET}, \textit{TH} and the following indicators:
{\setstretch{1.0}
\begin{itemize}
  \item \textbf{Propagation Delay} (\textit{PD}): the time required for a signal from an input pin to propagate through combinational logic and appear at an external output pin.
  \item \textbf{Look-Up Table} (\textit{LUT}): the number of combinational adaptive lookup tables required to implement an algorithm in hardware. The number of \textit{LUTs} is an indicator of the size of hardware in Altera devices. In other devices, the area could be measured in terms the total number of gates, logic elements, slices, etc.
  \item \textbf{Logic Register} (\textit{LR}): the total number of logic registers in the design.
  \item \textbf{Power Consumption} (\textit{PC)}: the power consumption of the developed hardware in Watts.
\end{itemize}}

\subsection{Outcomes}

The Outcomes element is the formulation of \textit{CMIs} as function of \textit{KIs}. The Lightness Indicator \textit{LI} is the main \textit{CMI} calculation in the presented statistical analysis framework. The \textit{LI} is calculated in terms of several \textit{AP}s; three for the current study, namely the General Algorithmic Profile (\textit{GAP}), Software Profile (\textit{SWP}), and Hardware Profile (\textit{HWP}). The simplified form of \textit{LI} is shown in Equation \ref{eqnLIRad}:

\begin{equation}
\label{eqnLIRad}
LI = \sqrt[l]{ratio_{1}\cdot ratio_{2} \cdot ratio_{3}...ratio_{l}}
\end{equation}
\noindent and hence

\begin{center}
$\textit{LI} = (\prod\limits_{i=1}^l \textit{ratio}_{i})^{\frac{1}{l}}$
\end{center}
\noindent Where $l$ is the number of key indicators.

The weighted version of \textit{LI} is denoted by \textit{wLI} in Equation \ref{eqnwLI}. The weighted version enables the emphasis of specific indicators. If all the assigned weights are equal, the \textit{wLI} is the same as \textit{LI}.

\begin{equation}
\label{eqnwLI}
\textit{wLI} = (\prod\limits_{k=1}^l \textit{ratio}_{k}^{w_{k}})^{\frac{1}{\sum\limits_{k=1}^l \textit{w}_{k}}}
\end{equation}

\noindent Where $w_{k}$ is the weight of the $k^{th}$ ratio.

The \textit{LIS} enables the classification of cryptographic algorithms according to their lightness. A higher \textit{LI} is achieved through a higher throughput, a more efficient memory performance, more compact size, less complexity, less power consumption, and less resource utilization. The \textit{LI} is either directly or inversely proportional to the indicators. The aim of the chosen proportion is to emphasize lightness; the proportions could be modified to capture other properties. The master \textit{LIS} formula using the developed indicators is shown in Equation \ref{eqnLIEX}. The indicators that are common to the Software (sw) and Hardware (hw) profiles are labeled with the profile name.


\begin{equation} \label{eqnLIEX}
\begin{split}
\small LI = \sqrt[14]{GAP \cdot SWP \cdot HWP}\\
\small GAP = \frac{AC_{ref}}{AC} \cdot \frac{KS_{ref}}{KS} \cdot \frac{NR_{ref}}{NR} \cdot \frac{BS_{ref}}{BS}\\
\small SWP = \frac{ET_{sw,ref}}{ET_{sw}} \cdot \frac{TH_{sw}}{TH_{sw,ref}} \cdot \frac{CPI_{ref}}{CPI} \cdot \frac{CMR_{ref}}{CMR}\\
\small HWP = \frac{ET_{hw,ref}}{ET_{hw}} \cdot \frac{TH_{hw}}{TH_{hw,ref}} \cdot \frac{PD_{ref}}{PD} \cdot \frac{LUT_{ref}}{LUT} \cdot \frac{LR_{ref}}{LR} \cdot \frac{PC_{ref}}{PC}
\end{split}
\end{equation}

The \textit{LIS} provides the following set of combined statistical indicators:

\noindent Complexity Indicator (\textit{CI}):

\noindent \small{$CI = \sqrt[5]{\frac{AC}{AC_{ref}} \cdot \frac{KS}{KS_{ref}} \cdot \frac{NR}{NR_{ref}} \cdot \frac{BS}{BS_{ref}} \cdot \frac{CPI}{CPI_{ref}}}$}

\noindent Security Strength Indicator (\textit{SSI}):

\noindent \small{$SSI = \sqrt[4]{\frac{AC}{AC_{ref}} \cdot \frac{KS}{KS_{ref}} \cdot \frac{NR}{NR_{ref}} \cdot \frac{BS}{BS_{ref}}}$}

\noindent Hardware Lightness Indicator (\textit{HLI}):

\noindent \small{$HLI = \sqrt[6]{\frac{ET_{hw,ref}}{ET_{hw}} \cdot \frac{TH_{hw}}{TH_{hw,ref}} \cdot \frac{PD_{ref}}{PD} \cdot \frac{LUT_{ref}}{LUT} \cdot \frac{LR_{ref}}{LR} \cdot \frac{PC_{ref}}{PC}}$}

\noindent Software Lightness Indicator (\textit{SLI}):

\noindent \small{$SLI = \sqrt[4]{\frac{ET_{sw,ref}}{ET_{sw}} \cdot \frac{TH_{sw}}{TH_{sw,ref}} \cdot \frac{CPI_{ref}}{CPI} \cdot \frac{CMR_{ref}}{CMR}}$}

\noindent Speed Indicator (\textit{SI}):

\noindent \small{$SI = \sqrt[5]{\frac{ET_{sw,ref}}{ET_{sw}} \cdot \frac{TH_{sw}}{TH_{sw,ref}} \cdot \frac{ET_{hw,ref}}{ET_{hw}} \cdot \frac{TH_{hw}}{TH_{hw,ref}} \cdot \frac{PD_{ref}}{PD}}$}

%
%
%
%
%
%
%
%
%
%
%
%
%
%

\subsection{Performance}

\begin{figure}[t]
   \caption{The six elements diagram of the \textit{LIS}}
   \label{fig:LISGBModel}
   \centering
        \includegraphics[width=0.8\textwidth]{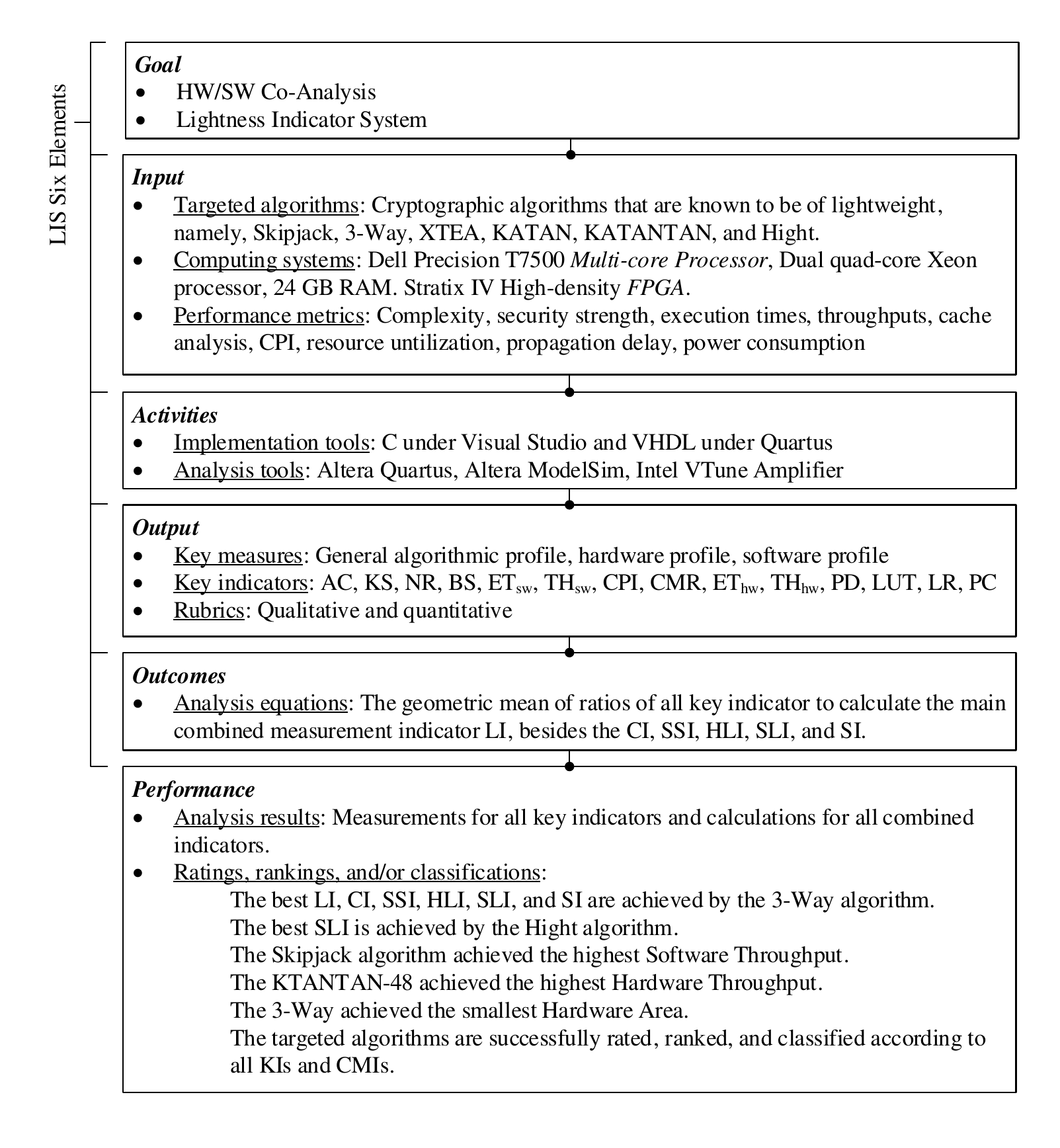}
\end{figure}

The analysis based on the \textit{LIS} \textit{Output} and \textit{Outcomes} provides measurements for all \textit{KIs} and enables the calculation of the defined \textit{CMIs}. The results include rating, ranking, and classifying the targeted algorithms. The analysis and evaluation of results are presented in Section~\ref{Performance Analysis and Evaluation}. The six elements of the \textit{LIS} are summarized in Figure~\ref{fig:LISGBModel}

\subsection{Programming Interface}
The developed statistical framework is embedded in a sample co-design \textit{IDE}. The \textit{IDE} is implemented using \textit{Java} under \textit{Netbeans}. Moreover, the code editor is implemented using \textit{RSyntaxTextArea} Java framework, while the \textit{IDE} theme is implemented using \textit{JTattoo} Java framework. The used implementation and performance evaluation tools comprise \textit{Altera} \textit{Quartus} and \textit{Altera ModelSim} for Hardware implementation and analysis, and In\textit{tel vTune Amplifier} under \textit{Visual Studio} for Software analysis. The developed IDE connects to \textit{Altera Quartus} using \textit{TCL} commands to synthesize and generate timing analyses, pin assignments for \textit{FPGA} boards, and generate bit files to program the targeted \textit{FPGAs}. The \textit{IDE} connects to \textit{Intel vTune Amplifier}, using Command Line and Batch Files, to perform the software analysis and calculating the total execution time, CPI, etc. The generated hardware and software analysis files are exported to \textit{MS Excel} to produce the complete analysis profile and charts.

\section{Analysis and Evaluation} \label{Performance Analysis and Evaluation}

\subsection{Performance Analysis} 
The \textit{LIS} is an application of an analysis framework that can be the core part of a benchmark within HW/SW co-design. The developed framework is applied by developing the \textit{LIS} on several cryptographic ciphers that are presented in the literature as lightweight, tiny, small, or minute. The developed system  enables the validation of the lightness of the algorithms through measurements and statistical analysis.

In accordance with our generic benchmark model, and upon the identification of the system \textit{Goal} and \textit{Input}, the \textit{Activities} are done according to the following procedure:

{\setstretch{1.0}
\begin{enumerate}
  \item Implement hardware using VHDL under \textit{Quartus}
  \item Analyze the hardware profile using \textit{Quartus} and \textit{ModelSim}
  \item Implement software using C-language under Visual Studio and its integrated \textit{Intel VTune Amplifier}
  \item Analyze the software profile using \textit{Intel VTune Amplifier}
  \item Derive and analyze the general algorithmic profile
  \item Combine and analyze the results from all profiles using a statistical software tool
  \newcounter{enumTemp}
    \setcounter{enumTemp}{\theenumi}
\end{enumerate}}

With the finalization of \textit{Activities}, the following steps complete the elements of the framework:

{\setstretch{1.0}
\begin{enumerate}
\setcounter{enumi}{\theenumTemp}
  \item Produce the \textit{Output} key indicators
  \item Calculate the combined indicators of the \textit{Outcomes}
  \item Build the overall \textit{Performance} report
\end{enumerate}}

Tables \ref{GAProfile}, \ref{SWProfile}, and \ref{HWProfile} present the derivation and implementation results of the general algorithmic, software, and hardware profiles. On the simple indicators level, the \textit{Skipjack} algorithm achieved the highest software execution throughput of $156.098 Mbps$, while the highest hardware execution throughput, $480 Mpbs$ is achieved by the \textit{KTANTAN-48} algorithm. The \textit{3-Way} algorithm attained the smallest hardware area of $77 ALUTs$ and $167 LRs$.

\begin {table}[h]
\small
\caption{General Algorithmic Profile} \label{GAProfile}
\begin{center}
\small
\begin{tabular}{c c c c c c}

\hline \textbf{Algorithm Name} & \textbf{AC} & \textbf{Mapped AC} & \textbf{KS} & \textbf{NR}& \textbf{BS}\\
\hline

Skipjack & AQ & 0.8 & 80 & 32 & 64\\
XTEA & AQ & 0.8 & 96 & 64 & 64\\
3-WAY &AQ & 0.8 & 128 & 11 & 96\\
HIGHT&AQ&0.8&128&32&64\\
KATAN-32&AQ&0.8&80&254&32\\
KATAN-48&AQ&0.8&80&254&48\\
KATAN-64&AQ&0.8&80&254&64\\
KTANTAN-32&AQ&0.8&80&254&32\\
KTANTAN-48&AQ&0.8&80&254&48 \\
KTANTAN-64&AQ&0.8&80&254&64 \\
\textbf{AES} &\textbf{AC}&\textbf{0.8}&\textbf{192}&\textbf{12}&\textbf{128}\\
\hline
\end{tabular}
\end{center}
\end{table}

\begin{table*}[h]
\caption{Software Profile} \label{SWProfile}
\begin{center}
\small
\begin{tabular}{c c c c c c}
\hline \textbf{Algorithm Name} & \textbf{BS} & \textbf{ET($\mu$sec)} & \textbf{TH(Mbps)} & \textbf{CPI}& \textbf{CMR}\\
\hline
Skipjack & 64.000 & 0.410 &156.098 & 1.327 &0.164\\
XTEA & 64.000&2.570&24.903&0.729&0.033\\
3-WAY &96.000&2.320&41.379&1.107&0.036\\
HIGHT&64.000&8.640&7.407&1.330&0.000\\
KATAN-32&32.000&27.460&1.165&0.634&0.006\\
KATAN-48&48.000&40.330&1.190&0.634&0.004\\
KATAN-64&64.000&52.830&1.211&0.627& 0.003\\
KTANTAN-32&32.000&791.080&0.040&0.986& 0.001\\
KTANTAN-48&48.000&803.320&0.060&0.975& 0.001\\
KTANTAN-64&64.000&821.830&0.078&0.965& 0.001\\
\textbf{AES} &\textbf{128.000}&\textbf{23.210}&\textbf{5.515}&\textbf{1.235}&\textbf{0.004}\\
\hline
\end{tabular}
\end{center}
\end{table*}

\begin{table*}[h]
\caption{Hardware Profile} \label{HWProfile}
\begin{center}
\small
\begin{tabular}{c p{1cm} p{1cm} p{1cm} c c c}
\hline \textbf{Algorithm Name} & \textbf{\small{ET}\tiny{($\mu$sec)} } & \textbf{\small{TH}\tiny{(Mpbs)} } & \textbf{\small{PD}\tiny{(nsec)} } & \textbf{\small{ALUT}} & \textbf{\small{LR}}& \textbf{\small{PC}\tiny{(mW)}}\\
\hline
Skipjack & 7.49 & 8.55 &11.90 &554.00&142.00&331.01\\
XTEA & 6.18 &10.35&11.10&2799.00&135.00&332.77\\
3-WAY &0.80&120.00&3.82&77.00&167.00&331.01\\
HIGHT&1.85&34.59&127.78&2036.00&72.00&332.66\\
KATAN-32&1.47&21.77&43.57&2145.00&540.00&328.63\\
KATAN-48&1.89&25.40&79.94&3982.00&556.00&329.95\\
KATAN-64&2.38&26.89&78.31&4315.00&572.00&330.94\\
KTANTAN-32&0.09&372.09&40.03&1947.00&112.00&328.58\\
KTANTAN-48&0.10&480.00&72.78&3662.00&128.00&329.81\\
KTANTAN-64&0.15&438.36&79.30&4075.00&144.00&331.00\\
\textbf{AES} &\textbf{1.46}&\textbf{6.83}&\textbf{5.35}&\textbf{3998.00}&\textbf{750.00}& \textbf{654.87}\\
\hline
\end{tabular}
\end{center}
\end{table*}



\begin{figure}[h]
   \caption{The Lightness Indicator classification}\label{LI}
   \centering
     \includegraphics[width=1.1\textwidth]{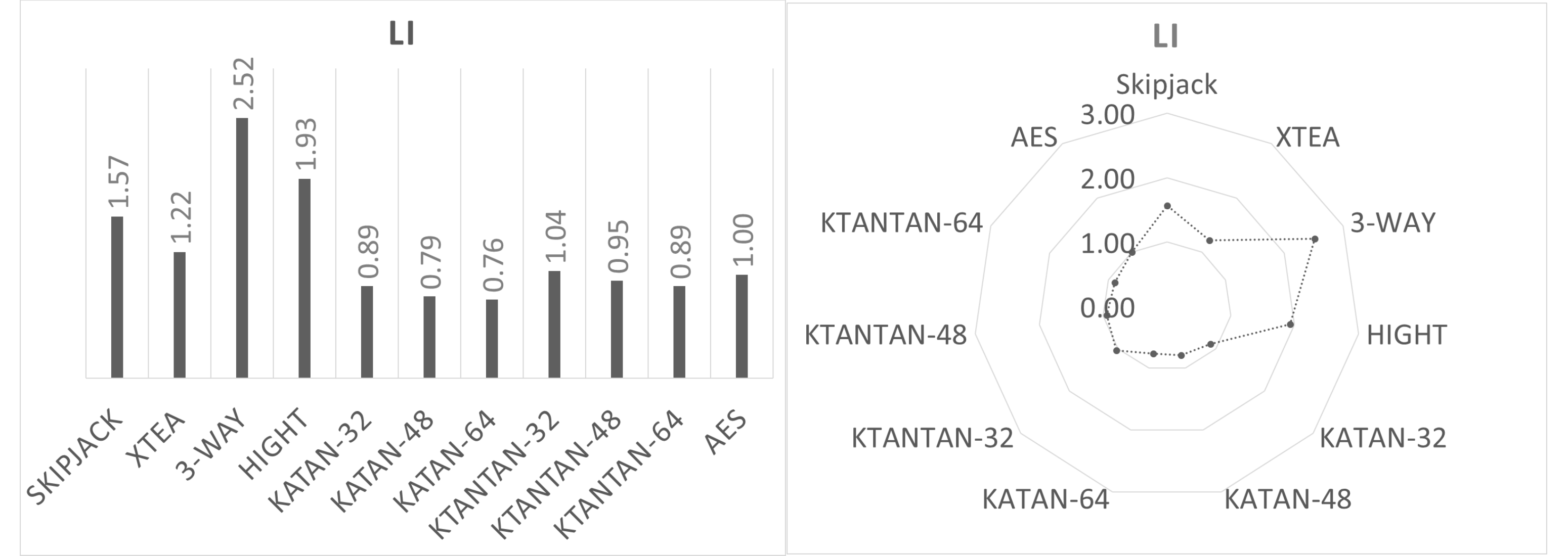} 
\end{figure}
\begin{figure}[h]
   \caption{The Software and Hardware Lightness Indicator classifications} \label{HSLI}
   \centering
     \includegraphics[width=1.1\textwidth]{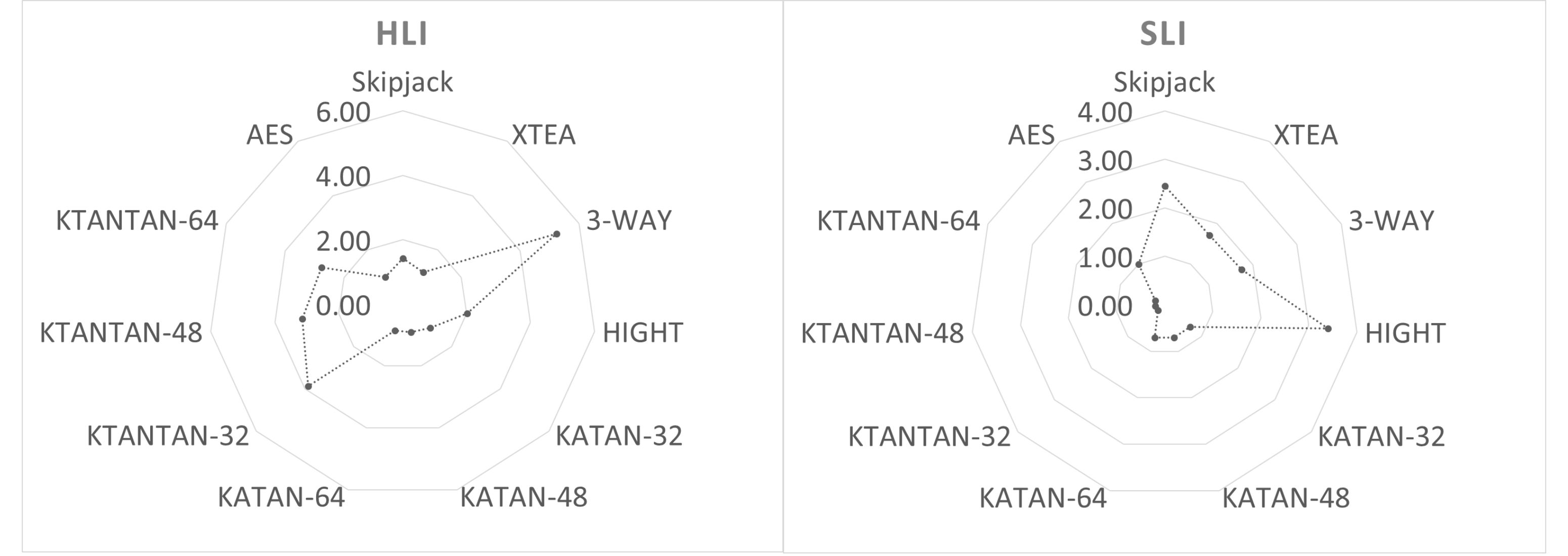}
\end{figure}
\begin{figure}[h]
	\caption{The Complexity and Security Strength Indicator classifications}  \label{CSI}
	\centering
	\includegraphics[width=1.1\textwidth]{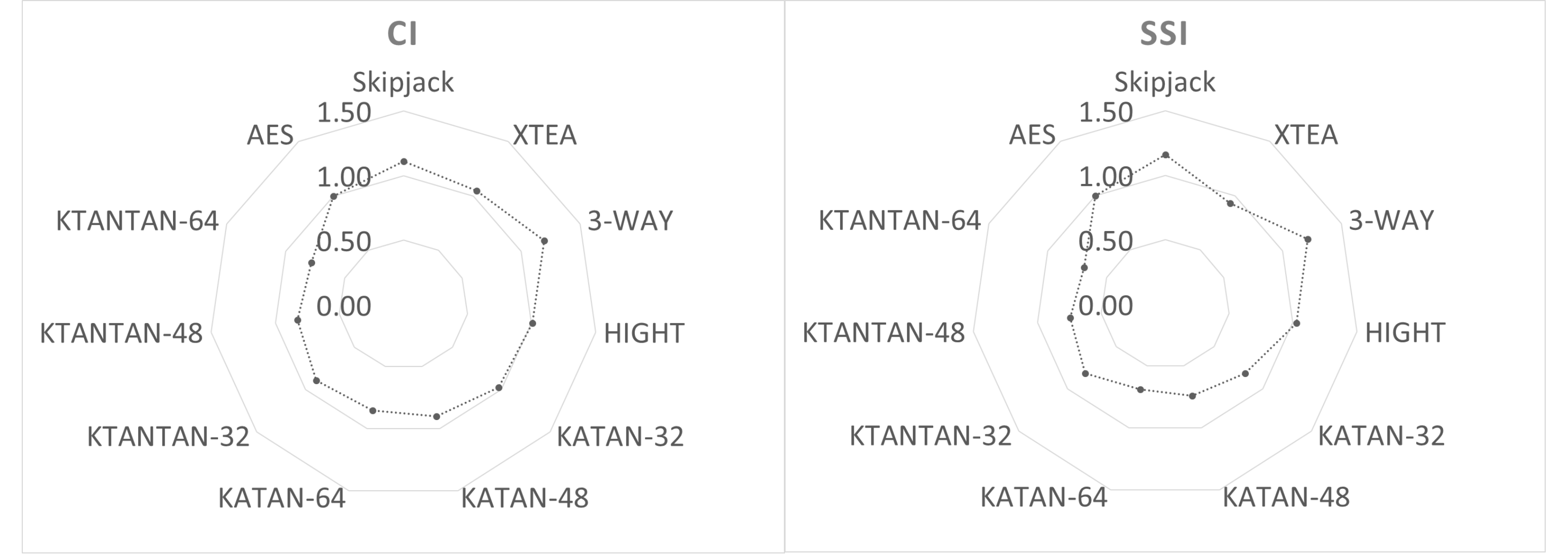}
\end{figure}
\begin{figure}[h]
   \caption{The Speed Indicator classification} \label{SI}
   \centering
     \includegraphics[width=0.65\textwidth]{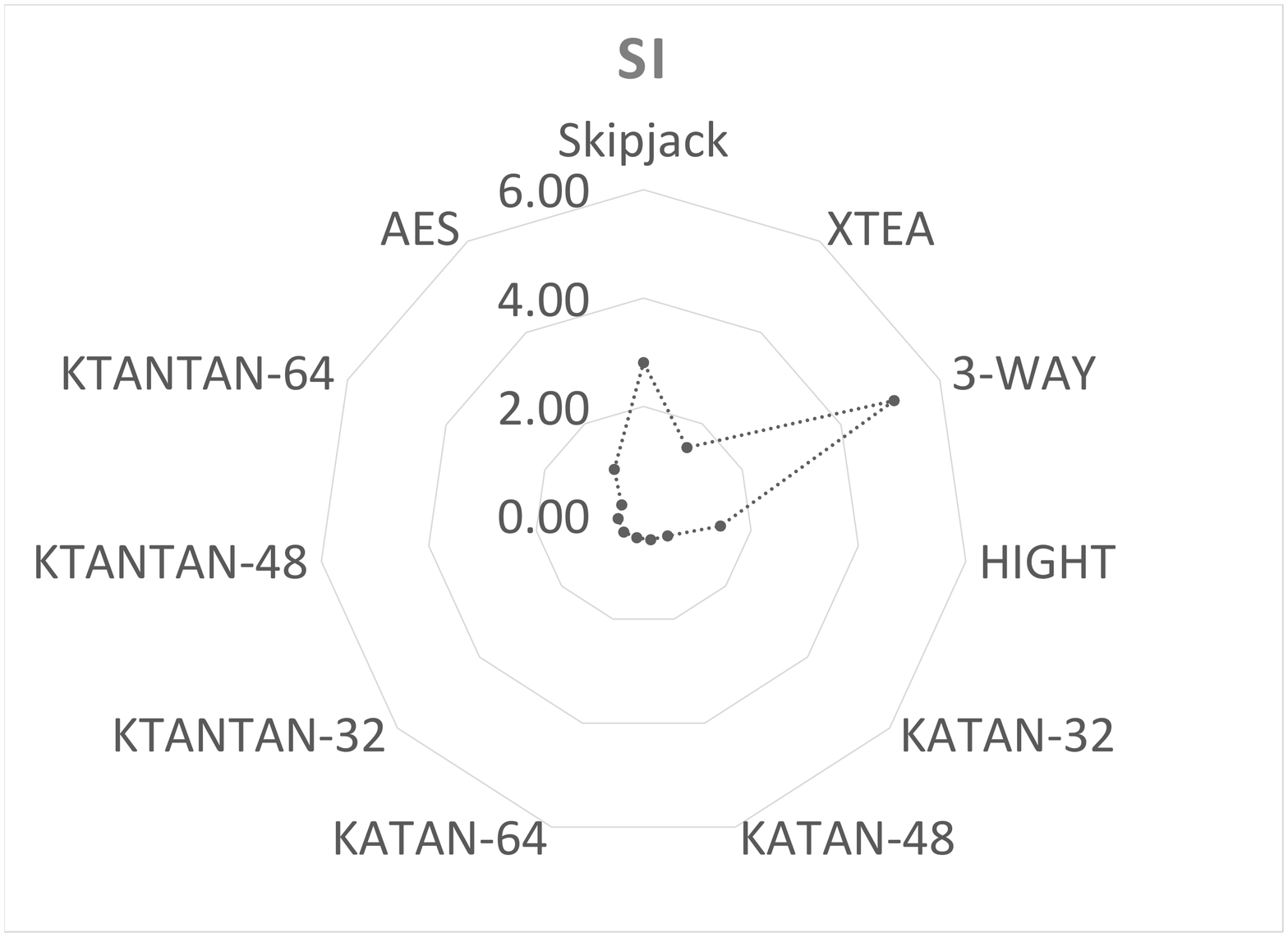}
\end{figure}

The \textit{Lightness}, \textit{Complexity}, \textit{Security Strength}, \textit{Hardware Lightness}, \textit{Software Lightness}, and \textit{Speed} indicators are shown in Table \ref{Indicators}. The algorithm that attained a larger indicator value is lighter, of a less complexity, of a higher security strength, or faster than the algorithm with a lower indicator value. Figures \ref{LI}, \ref{HSLI}, \ref{CSI}, and \ref{SI} present a per-\textit{CMI} comparison. The \textit{3-Way} got the best lightness index of $2.52$, while $KATAN-64$ attained the lowest with an index of $0.76$. The best \textit{CI} index is $1.19$; attained by the \textit{3-Way} algorithm. The best \textit{SSI}, \textit{HLI}, and \textit{SI} indices are $1.16$, $5.24$, and $5.08$; all attained by the \textit{3-Way} algorithm. The HIGHT algorithm achieved the highest \textit{SLI} index of $3.4$.

\begin{table}[h]
\caption{Indicators} \label{Indicators}
\begin{center}
\small
\begin{tabular}{c c c c c c c}
\hline \textbf{Algorithm Name} & \textbf{LI} & \textbf{CI} & \textbf{SSI} & \textbf{HLI}& \textbf{SLI}& \textbf{SI}\\
\hline
\textbf{Skipjack} &1.57&1.11&1.16&1.42&2.46&2.81 \\
\textbf{XTEA} &1.22&1.05&0.93&1.18&1.70&1.48\\
\textbf{3-WAY} &2.52&1.19&1.22&5.24&1.75&5.08\\
\textbf{HIGHT}&1.93&1.01&1.03&2.02&3.40&1.43\\
\textbf{KATAN-32}&0.89&0.98&0.82&1.12&0.69&0.59\\
\textbf{KATAN-48}&0.79&0.90&0.74&0.90&0.70&0.47\\
\textbf{KATAN-64}&0.76&0.85&0.69&0.86&0.71&0.44\\
\textbf{KTANTAN-32}&1.04&0.89&0.82&3.88&0.18&0.48\\
\textbf{KTANTAN-48}&0.95&0.83&0.74&3.14&0.20&0.47 \\
\textbf{KTANTAN-64}&0.89&0.78&0.69&2.76&0.21&0.44\\
\textbf{AES} &1.00&1.00&1.00&1.00&1.00&1.00\\
\hline
\end{tabular}
\end{center}
\end{table}

%

\subsection{General Evaluation} 

The current investigation can be evaluated at the levels of the framework development, application, and contextualization. The developed framework is unique in combining algorithmic, hardware, and software characteristics to provide unified performance evaluation criteria and useful performance indicators. The framework addresses the need for methods that can analyze the performance and deal with the hybrid nature of modern computing systems. The investigation proposes the creation of unified indexes/indicators that can capture specific qualities in terms of a wide range of heterogeneous key performance indicators, such as the \textit{LIS}. The \textit{LI} served as a master \textit{CMI} while an indicator like the \textit{SI} is developed with focus on speed. Indeed, the framework is scalable and upgradeable without changing the statistical computation or the structure of the measurement. For instance, an additional \textit{AP} can be incorporated into the calculations of the \textit{LIS} to include the performance characteristics of \textit{GPUs}. 

At the application level, the developed framework can be used to examine qualities of importance and interest to developers or users. For example, the presented \textit{LIS} enables the indexing and classifying of cryptographic algorithms. Here, the qualities of importance are the lightness, speed, complexity, and security strength. The \textit{LIS} can be applied to examine the same qualities for a similar area of application, such as, signal processing. In signal processing, the \textit{SLI} and \textit{HLI} can be reused. However, the \textit{LI} and \textit{CI} needs to be redefined within the context of signal processing, and the \textit{SSI} is not applicable. Signal processors are usually embedded within real-time application and characterized by their numerical accuracy, acceleration schemes, and the ability to perform fast computations and data access  \cite{ingle2016digital}. A \textit{Reliability and Accuracy Indicator} (\textit{RAI}) \textit{CMI} can be created to combine the desired characteristics and capture, index, and aid in the classification of signal processing algorithms. 

The contextualization of the framework in relation to the targeted area of application produces a rich and comprehensive set of reference \textit{KIs}. \textit{KIs}, such as \textit{ET}, \textit{TH}, \textit{CPI}, \textit{CMR}, \textit{PD}, \textit{LUT}, \textit{LR}, and \textit{PC}, are independent of the context of application and thus highly reusable. Other \textit{KIs}, such as \textit{KS}, \textit{NR}, and \textit{BS} are specific to cryptographic algorithms. \textit{KIs} can measure quantities or describe qualities. Qualities can be easily specified using rubrics and mapped onto quantities that can be substituted into the indicator equations. \textit{KIs} should be carefully identified and developed by experts in the targeted area of application, and supported by evident applicability. The contextualization of the application for the proposed \textit{RAI}, in signal processing, can comprise \textit{KIs}, such as, \textit{Memory Access Time} as a quantitative measurement. The availability of \textit{Specialized Addressing Modes} can be captured as a qualitative indicator.

This paper presents a generic model that specifies the elements of benchmarks and/or analysis frameworks. The benchmark model is used to present the \textit{LIS}, nevertheless the model can be used to describe any benchmark. The developed model is generic, simple, concise, and aids the clear description of benchmarks using a unified pattern.   

The developed statistical framework is applied through a case-study that targets a class of cryptographic algorithms. The selected algorithms are presented in the literature as tiny, small, minute, and light. The case-study provided a unified classification criteria that include \textit{LI}. The proposed framework successfully classified the targeted algorithms according to their hardware, software, and algorithmic characteristics. The addressed algorithms are widely implemented, analyzed, and evaluated in the literature. The work presented in the literature is limited to single algorithm evaluation, single system implementation, such as either hardware or software, and still make holistic claims of lightness based on limited number of indicators. The used reference implementation is the \textit{AES} cipher with a key size of $192$ bits. The use of other key sizes, such as $256$ bits, doesn't change the algorithm classifications or falsify the analysis as it is consistently applied for all the targeted algorithms. However, different indicator values are expected.

The developed framework is intended to capture hardware and software properties. The current investigation is limited to non-partitioned implementations, where the whole computation is delegated to a co-processor. Partitioned implementations can be evaluated, based on the proposed framework, by analyzing the \textit{KIs} of the hardware and software subsystems. The obtained \textit{KI} measurements capture the subsystem characteristics. In addition, carefully defined \textit{CMIs} can rank, rate, and classify different partitioning strategies per optimization target, such as, area, speed, power consumption, etc. 

An example similar investigation is presented by Spacey et al. in \cite{Spacey2009159}. The authors combined several hardware and software characteristics within a heuristic to produce a single execution time estimate. The best time estimate is identified based on heterogeneous performance and architectural characteristics of different hardware and software partitions. Our proposed framework would, with no doubt, enrich such investigations and provide versatile estimates with \textit{CMIs} such as \textit{HLI}, \textit{SLI}, \textit{CI}, \textit{SI}, and/or other customized indicators.

\section{Conclusion} \label{Conclusion}

Modern high-performance computers are hybrids of multi-core processors, \textit{GPUs}, \textit{FPGAs}, etc. In this paper, a statistical framework is developed to provide thorough analysis and evaluation of algorithms and their implementations on different processing systems. A generic benchmark model is created to present the framework with clarity. The framework categorizes processing subsystems into profiles, where each can be contextualized according to a specific application. The statistical framework is adopted to analyze and evaluate a set of cryptographic algorithms that are claimed to be small in size, tiny, and efficient. The proposed framework enabled the creation of several key indicators including the lightness, complexity, security strength, and speed indicators.  The two main targeted high-performance computing devices are multi-core processors for software implementations and high-end \textit{FPGAs} for hardware implementations. The developed lightness indicator ranks the \textit{3-Way} algorithm as the lightest among all with an \textit{LI} of 2.52. \textit{Hight} achieves the second best lightness with a score of 1.93. The lowest score of 0.79 was attained by \textit{KATAN-64}. The case-study validates the statistical framework and leads to a successful classification of the targeted algorithms. The obtained results are based on a combination of three profiles including the algorithmic, software, and hardware profiles. The presented framework enjoys being scalable, upgradeable, and portable across-applications. Future work includes incorporating additional processing systems, targeting other areas of application, and embedding the framework within a co-design \textit{IDE} and target partitioned implementations. 

\section*{References}

\bibliography{ref}

\section*{Author biographies}
\noindent {\bf Issam Damaj, PhD ME BE,} is an Associate Professor of Computer Engineering at the American University of Kuwait. His research interests include hardware/software co-design, embedded system design, automation, Internet-of-things, and engineering education. He is a Senior Member of the IEEE and a Professional Member of the ASEE. He maintains an academic website at www.idamaj.net.\\

\noindent {\bf Safaa Kasbah, MSc BSc, } received a Master Degree in Computer Science, in 2006, from the Lebanese American University. She received a Bachelor Degree in Computer Science and a minor in Physics from the American University of Beirut in 2004. Her main research interests are iterative methods, hardware/software co-design, reconfigurable computing, quantum computing and Information and Knowledge Management.

\end{document}